\documentclass[11pt]{article}

\usepackage{amsmath,amsfonts,amsthm}
\usepackage{graphicx}
\usepackage{fixmath}
\usepackage{amssymb,latexsym}
\usepackage{mathrsfs,amsbsy}
\usepackage{dsfont}
\usepackage{algorithm}
\usepackage{eucal}
\usepackage{colortbl}
\usepackage{multirow}
\usepackage{float}
\graphicspath{{./figures/}}
\usepackage{booktabs}
\usepackage{caption}
\usepackage{subfigure}
\usepackage{microtype}
\usepackage{enumerate}
\usepackage{hyperref}

\usepackage[margin=0.5in]{geometry}
\usepackage{tikz} 
\usetikzlibrary{shapes,arrows} 

\theoremstyle{definition}

\theoremstyle{remark}

\numberwithin{equation}{section}
\numberwithin{subsection}{section}

\title{Computational Technologies for Brain Morphometry}
\author{Zicong Zhou\thanks{University of Texas at Arlington, Mathematics Department, Arlington, Texas 76019-0408, zicong.zhou@mavs.uta.edu}
\and Ben Hildebrandt\thanks{University of Texas at Arlington, Mathematics Department, Arlington, Texas 76019-0408, ben.hildebrandt@mavs.uta.edu}
\and Xi Chen\thanks{University of Texas at Arlington, Mathematics Department, Arlington, Texas 76019-0408, xi.chen@mavs.uta.edu}
\and Guojun Liao\thanks{University of Texas at Arlington, Mathematics Department, Arlington, Texas 76019-0408, liao@uta.edu}
\date{October 10, 2018}
}

\begin{document}
\maketitle    
\begin{abstract}
 In this paper, we described a set of computational technologies for image analysis with applications in Brain Morphometry. The proposed technologies are based on a new Variational Principle which constructs a transformation with prescribed Jacobian determinant (which models local size changes) and prescribed curl-vector (which models local rotations). The goal of this research is to convince the image research community that Jacobian determinant as well as curl-vector should be used in all steps of image analysis. Specifically, we develop an optimal control method for non-rigid registration; a new concept and construction of average transformation; and a general robust method for construction of unbiased template from a set of images. Computational examples are presented to show the effects of curl-vector and the effectiveness of optimal control methods for non-rigid registration and our method for construction of unbiased template.

\end{abstract}
\section{Introduction}
In this paper, we address a key mathematical issue in brain morphometry which investigates variabilities in the shape and size of brain structures from image data. The issue is how to characterize diffeomorphisms. The prevailing paradigm in this field is to focus on Jacobian determinant $J(\pmb{T})$ of the transformation $\pmb{T}$, which models local size changes. We show that $J(\pmb{T})$ alone cannot completely determine a transformation. Instead, we must use both $J(\pmb{T})$ and the curl-vector of $\pmb{T}$, $\text{curl}(\pmb{T})$, which models local rotaions, in all steps of brain morphometry studies. Specifically, we propose a new set of computational tools for morphometry studies. 
\begin{enumerate}[(1)]
	\item For the non-rigid registration step, we develop an optimal control method which minimizes a dis-similarity measure ($DS$) under the constraints of partial differential equations such as $\text{div}(\pmb{T}) =f$, $\text{curl}(\pmb{T})=g$ with respect to control functions $f$ and $g$.
	
	\item For the construction of an unbiased template, we propose an innovative concept of averaging a set of diffeomorphisms based on averaging the Jacobian determinants and curl-vectors of the diffeomorphisms.
	 
	\item For group differences, we propose to use Jacobian determinants and curl-vectors as features in TBM (tensor based morphometry) studies.
\end{enumerate} 

In this paper, we will focus on (1) and (2). We will demonstrate the effectiveness of our optimal control registration method mentioned in (1) and use it to perform the needed registration for constructing an unbiased template from a set of images in (2). In a future work, we will implement (3) in TBM (tensor based morphometry) studies of real data sets.

\section{Effect of curl-Vectors, Variational Principle, Average of Transformations}
The computational technologies described in this paper are based on a newly formulated Variational Principle \cite{XiChen2}. We begin with an example to show the effects of the curl-vector.
\subsection{Effect of curl-Vectors}
Started with a ground truth image (GT), based on portrait of Mona Lisa with 128*128 pixels shown in Figure 1(a). Then we generate two deformations $\pmb{D}_{\pmb{1}}$ and $\pmb{D}_{\pmb{2}}$ from the 128*128 uniform grid Figure 1(d) with spacing 1 (shown in 32*32) by opposite rotations such as that their Jacobian determinants are very close to 1 (in fact, between 0.996 and 1.003), and the average of their curl is 0 (the two rotations are cancelled). Figure 1(b) and 1(c) are the re-sampled images of Figure 1(a) on these two deformations Figure 1(e) and 1(f), respectively. 
\subsubsection*{Resampling $I_{0}$ by $\pmb{D}_{1}$ and $\pmb{D}_{2}$, which $J(\pmb{D}_{2}) \approxeq 1 \approxeq J(\pmb{D}_{1})$ but $\text{curl}(\pmb{D}_{1}) \ne \text{curl}(\pmb{D}_{2})$ }
\begin{figure}[H]
	\caption{}
	\subfigure[$I_{0}$ --- Ground Truth]{\includegraphics[width=4.7cm,height=4.7cm]{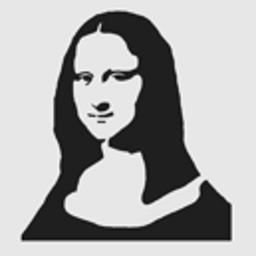}}
	\hspace{2.5cm}	
	\subfigure[$I_{1}$]{\includegraphics[width=4.7cm,height=4.7cm]{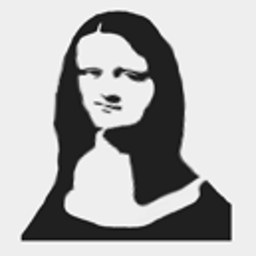}}
	\hspace{2.5cm}
	\subfigure[$I_{2}$]{\includegraphics[width=4.7cm,height=4.7cm]{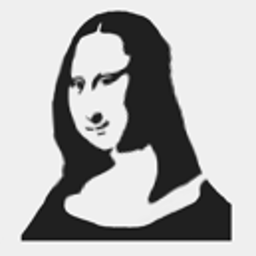}}
	
	\subfigure[$\pmb{Id}$ --- uniform grid]{\includegraphics[width=4.7cm,height=4.7cm]{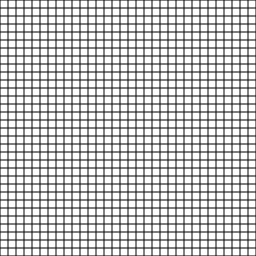}}
	\hspace{2.5cm}	
	\subfigure[$\pmb{D}_{1}$]{\includegraphics[width=4.7cm,height=4.7cm]{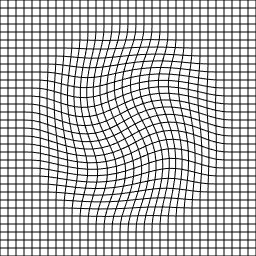}}
	\hspace{2.5cm}
	\subfigure[$\pmb{D}_{2}$]{\includegraphics[width=4.7cm,height=4.7cm]{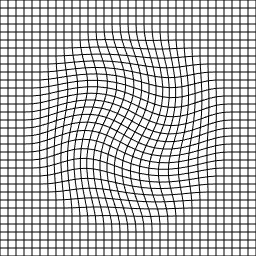}}
\end{figure}
 
\subsection{Variational Principle}

In this section, we describe the principles in 2D \cite{XiChen2}, where $\text{curl}(\pmb{T})$ is a scalar function. A diffeomorphism $\pmb{T}_1$ on a square domain $\mathbb{D}$ can be deformed to $\pmb{T}_2$ with prescribed Jacobian determinant $J(\pmb{T}_2(\pmb{x})) = f_0(\pmb{x})$ and $\text{curl}(\pmb{T}_2(\pmb{x})) = g_0(\pmb{x})$ in the interior of $\mathbb{D}$, and $\pmb{T}(\pmb{x})= \pmb{x}$ on the boundary of $\mathbb{D}$ ($f_0$ is properly normalized for solvability). $\pmb{T}_2$ is constructed by computational minimization of the functional:

\begin{equation}
\iint_{\mathbb{D}}{[(J(\pmb{T}(\pmb{x}))-f_0(\pmb{x}))^2+(\text{curl}(\pmb{T}(\pmb{x})-g_0(\pmb{x}))^2]}d\pmb{x}
\end{equation}

\noindent subject to the constraints $\mathrm{\Delta} \pmb{T}=\pmb{F}$. The variational problem is computed by a gradient decent method with respect to the control function $\pmb{F}= (F_1, F_2)$. We demonstrate the principle in next two examples below. 
\subsubsection*{Example 2: Numerical Performances of Variational Principal}
Deform $\pmb{D}_{1}$ to $\pmb{D}_{2}$ by the Variational Principle. Taking $f_0 = J(\pmb{D}_{2})$ and $g_0 = \text{curl}(\pmb{D}_{2})$ in the Variational Principle, in 25 iteration steps, $\pmb{D}_{1}$ is deformed to a transformation --- $\hat{\pmb{D}}_{12}$, as shown in Figure 2(f) that is almost identical to $\pmb{D}_{2}$ by prescribing its Jacobian as $J(\pmb{D}_{2})$ and its curl as $\text{curl}(\pmb{D}_{2})$. Total elapsed time for 25 iteration steps is 0.955899 seconds using a laptop.
\begin{figure}[H]
	\begin{center}
		\caption{$\pmb{D}_{1}$ to $\pmb{D}_{2}$}
		\subfigure[$\pmb{D}_{1}$]{\includegraphics[width=4.6cm,height=4.6cm]{MD1.png}}
		\includegraphics[width=4.6cm,height=4.6cm]{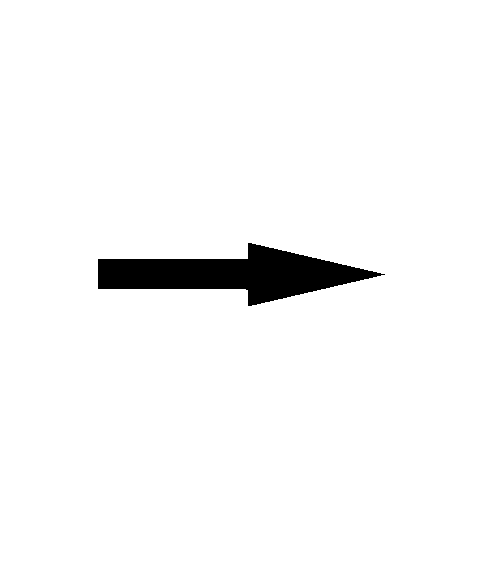}
		\subfigure[$\pmb{D}_{2}$]{\includegraphics[width=4.6cm,height=4.6cm]{MD2.png}}
	\end{center}
\end{figure}

\begin{figure}[H]
	\subfigure[step 0]{\includegraphics[width=4.6cm,height=4.6cm]{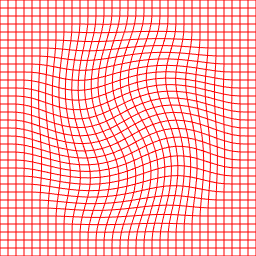}}
	\subfigure[step 2]{\includegraphics[width=4.6cm,height=4.6cm]{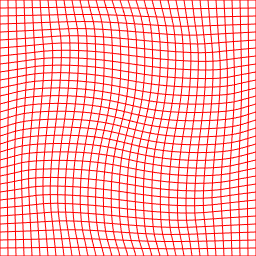}}
	\subfigure[step 10]{\includegraphics[width=4.6cm,height=4.6cm]{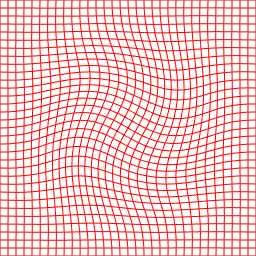}}
	\subfigure[$\hat{\pmb{D}}_{12}$ superimposed on $\pmb{D}_{2}$(black) ]{\includegraphics[width=4.6cm,height=4.6cm]{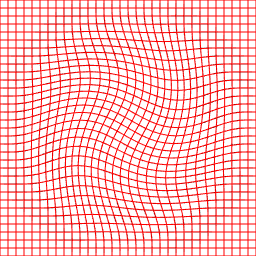}}
\end{figure}
We next add noise to $\pmb{D}_{1}$ and deform the distorted $\pmb{D}_{1}$, denoted as $\pmb{Dn}_{1}$, to $\pmb{D}_{2}$ as before. The calculated transformation $\hat{\pmb{Dn}}_{12}$ (in red lines) by The Variational Principle is superimposed on $\pmb{D}_{2}$ as it shows in Figure 3(f). The red lines are almost identical to the black lines. This indicates that the results of the Variational Principle are very accurate.
\begin{figure}[H]
	\begin{center}
		\caption{$\pmb{Dn}_{1}$ to $\pmb{D}_{2}$}
		\subfigure[$\pmb{Dn}_{1}$]{\includegraphics[width=4.6cm,height=4.6cm]{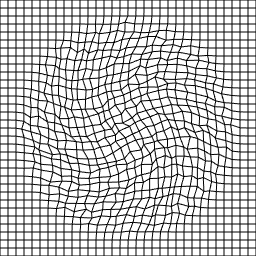}}
		\includegraphics[width=4.6cm,height=4.6cm]{Go2R.png}		\subfigure[$\pmb{D}_{2}$]{\includegraphics[width=4.6cm,height=4.6cm]{MD2.png}}
	\end{center}
\end{figure}
\begin{figure}[H]
	\subfigure[step 0]{\includegraphics[width=4.6cm,height=4.6cm]{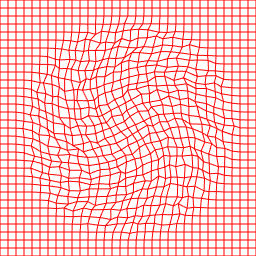}}
	\subfigure[step 550]{\includegraphics[width=4.6cm,height=4.6cm]{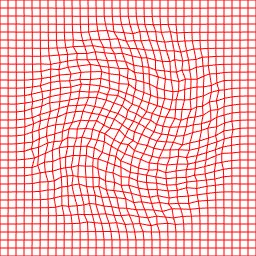}}
	\subfigure[step 7200]{\includegraphics[width=4.6cm,height=4.6cm]{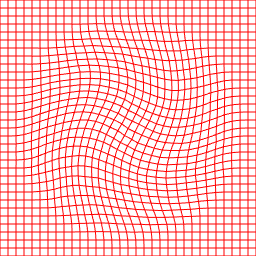}}
	\subfigure[$\hat{\pmb{Dn}}_{12}$ superimposes on $\pmb{D}_{2}$(black) ]{\includegraphics[width=4.6cm,height=4.6cm]{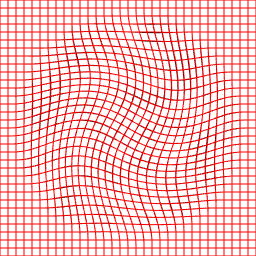}}
\end{figure}
\subsection{Average of Transformations}
Based on the Variational Principle, we propose to average a set of transformations by averaging their Jacobian determinants and curl vectors. 
Given a set of transformations $\pmb{T}_{i}$ where $i =1, 2, \dots, N$, we construct their average in the following steps: 
\begin{enumerate}[(1)]
	\item Let $f_0 = \frac{1}{N} \sum_{i}^{N}{J(\pmb{T}_{i})}$ and $g_0 = \frac{1}{N}\ \sum_{i=1}^{N}{\text{curl}(\pmb{T}_{i=1})}$ as in \cite{XiChen3}.  (weighted averages can also be used)
	\item Use $f_0$ and $g_0$ in the Variational Principle to calculate a transformation $\pmb{T}$ such that $J(\pmb{T}) = f_0$ and $\text{curl}(\pmb{T}) = g_0$. 
	\item We define this $\pmb{T}$ as the average of $\pmb{T}_i$, where $i= 1, \dots, N$. 
\end{enumerate}

Our definition of average deformation has clear geometrical meaning: The local size change ratios modeled by $J(\pmb{T}_i)$ and the local rotations modeled by $\text{curl}(\pmb{T}_i)$ are averaged to determine the average deformation. These two geometrical features also are biologically meaningful: the Jacobian determinant models the tissue size changes; the curl vector models the local shape change. Moreover, we directly work with transformations and thus the method applies to different registration methods.

\section{Construction of Unbiased Template}
We propose a new method for construction of unbiased template from the members of an image set. The general method is described in Example 3 and a competitive computation friendly approach in Example 4. First, we obtain six brain images by re-sampling a ground truth brain image on six intentionally designed transformations $\pmb{D}_i$, where $i =1,\dots,6$.

\begin{figure}[H]
	\caption{$I_{0}$ ground truth (GT) image}
	\begin{center}
		\includegraphics[width=3.1cm,height=3.1cm]{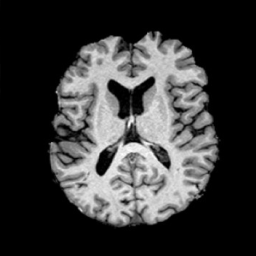}
	\end{center}
\end{figure}

We constructed six transformations $\pmb{D}_{i}$ where $i =1, \dots, 6$, which are shown below:
\begin{figure}[H]
	\caption{Transformations $\pmb{D}_{1-6}$}
	\subfigure[$\pmb{D}_{1}$]{\includegraphics[width=3.1cm,height=3.1cm]{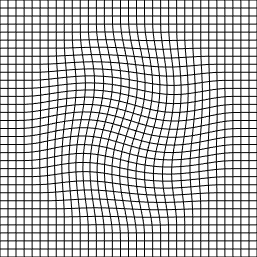}}
	\subfigure[$\pmb{D}_{2}$]{\includegraphics[width=3.1cm,height=3.1cm]{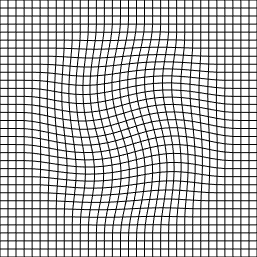}}
	\subfigure[$\pmb{D}_{3}$]{\includegraphics[width=3.1cm,height=3.1cm]{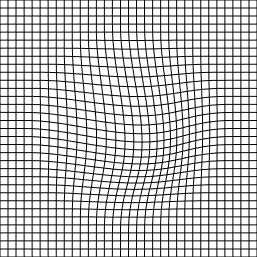}}
	\subfigure[$\pmb{D}_{4}$]{\includegraphics[width=3.1cm,height=3.1cm]{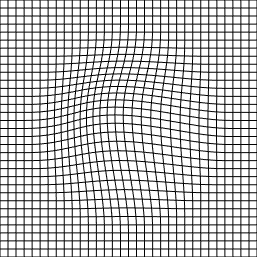}}
	\subfigure[$\pmb{D}_{5}$]{\includegraphics[width=3.1cm,height=3.1cm]{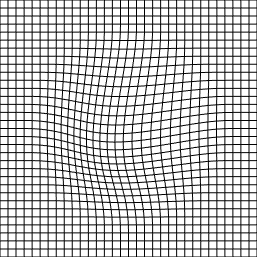}}
	\subfigure[$\pmb{D}_{6}$]{\includegraphics[width=3.1cm,height=3.1cm]{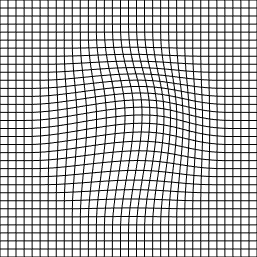}}
\end{figure}

\noindent These transformations are constructed in such a way that their Jacobian determinants have average equal to 1, and their curls have average 0. In fact, we have
 \begin{equation*}
	 \begin{aligned}
	0.999986099590103 &\le \frac{1}{6} \sum_{i=1}^{6}{J(\pmb{D}_i)} \le 1.000016719949570 \\
	-6.4029*(10^{-6}) &\le \frac{1}{6}\sum_{i=1}^{6}{\text{curl}(\pmb{D}_i)} \le\ 6.4029*(10^{-6})
	 \end{aligned}
 \end{equation*}

\noindent Six modeled brain images $I_i$, where $ i = 1,\dots,6$ are generated by re-sampling GT on each of the six transformations $\pmb{D}_i$, where $i =1,\dots,6$. These images are shown Figure 6. 
\begin{figure}[H]
	\caption{Image $I_{1-6}$}
	\subfigure[$I_{1}$]{\includegraphics[width=3.1cm,height=3.1cm]{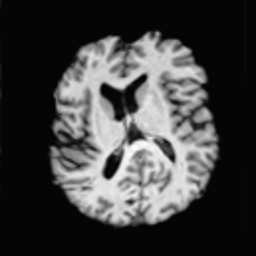}}
	\subfigure[$I_{2}$]{\includegraphics[width=3.1cm,height=3.1cm]{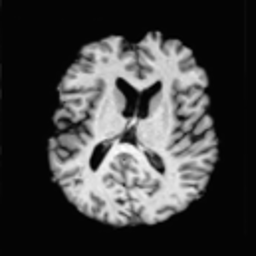}}
	\subfigure[$I_{3}$]{\includegraphics[width=3.1cm,height=3.1cm]{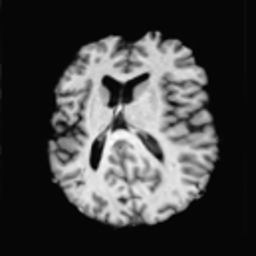}}
	\subfigure[$I_{4}$]{\includegraphics[width=3.1cm,height=3.1cm]{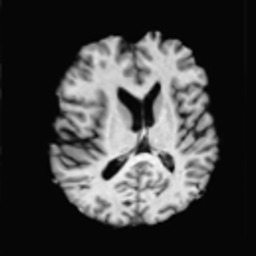}}
	\subfigure[$I_{5}$]{\includegraphics[width=3.1cm,height=3.1cm]{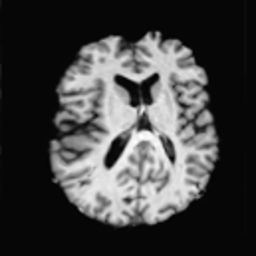}}
	\subfigure[$I_{6}$]{\includegraphics[width=3.1cm,height=3.1cm]{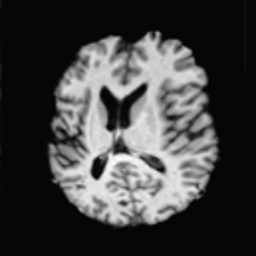}}
\end{figure}

\noindent Now when we take $f_0 = \frac{1}{6} \sum_{i}^{6}{J(\pmb{D}_i)} \approx 1$ and $g_0 = \frac{1}{6}\ \sum_{i}^{6}{\text{curl}(\pmb{D}_i)}\ \approx 0$, by the Variational Principle, the average of these images is expected to be a good approximation to GT. We will verify this in Example 3. The Sum of Squared Differences $(SSD)$ between $I_{i}$ and GT --- $SSD(I_{i}, I_{0})$ (they can be seen as the initial difference from $I_i$ to $I_0$) are shown in the following table:

\vspace{0.2cm}
\begin{minipage}{2\textwidth}
	\begin{tabular}{|c|c|c|c|c|c|c|}
		\hline
		$i$ & $1$ & $2$ & $3$ & $4$ & $5$ & $6$ \\
		\hline
		$SSD(I_i, I_0)=(10^{6})*$ & 1.8923 & 1.8906 & 1.8986 & 1.9610 & 1.8309 & 1.8353  \\
		\hline
	\end{tabular}
\end{minipage}
\vspace{0.2cm}
 
\subsubsection*{Example 3: General Approach}
In this example, we will walk through our method for the construction of unbiased template in a general sense. Step 1: take one image $I_i$ out of the six images as the initial template, then register $I_i$ to all six images $I_j$ for $j=1,2,\dots,6$ (arrowed dash-lines) to find six registration transformations --- $\pmb{\phi}_{ij}$ for $j=1,2,\dots,6$; Step 2: find the average transformation $\pmb{avg}_i=avg(\pmb{\phi}_{ij=1,\dots,6})$ of the six registration deformations by the Variational Principle (arrowed solid-line), as are demonstrated in the following diagram. 

\tikzstyle{block} = [rectangle, draw, fill=gray!20, text width=2cm, text centered, rounded corners, minimum height=1cm, node distance=2.5cm]
\tikzstyle{cloud} = [ellipse, fill=red!10, node distance=2.5cm, minimum height=2em]
\begin{center}
	\begin{tikzpicture}
	\node [block] (I1) {$I_{1}$};
	\node [block, right of=I1] (I2) {$I_{2}$};
	\node [block, right of=I2] (I3) {$I_{3}$};
	\node [block, right of=I3] (I4) {$I_{4}$};
	\node [block, right of=I4] (I5) {$I_{5}$};
	\node [block, right of=I5] (I6) {$I_{6}$};
	
	\node [cloud, below of=I1] (phi1) {$\pmb{\phi}_{i1}$};
	\node [cloud, right of=phi1, below of=I1] (phi2) {$\pmb{\phi}_{i2}$};
	\node [cloud, right of=phi2, below of=I2] (phi3) {$\pmb{\phi}_{i3}$};
	\node [cloud, right of=phi3, below of=I3] (phi4) {$\pmb{\phi}_{i4}$};
	\node [cloud, right of=phi4, below of=I4] (phi5) {$\pmb{\phi}_{i5}$};
	\node [cloud, right of=phi5, below of=I5] (phi6) {$\pmb{\phi}_{i6}$};
	\node [cloud, right of=phi6, below of=I6] (avgphi) {$\pmb{avg}_i$};
	
	\node [block, below of=phi3] (iTi) {$I_{i}$ (initial template)};

	\draw [-latex', dashed, bend left=10] (phi1) to (I1);
	\draw [-latex', dashed, bend left=6] (phi2) to (I2);
	\draw [-latex', dashed] (phi3) to (I3);
	\draw [-latex', dashed, bend right=3] (phi4) to (I4);
	\draw [-latex', dashed, bend right=5] (phi5) to (I5);
	\draw [-latex', dashed, bend right=8] (phi6) to (I6);
	
	\draw [-, dashed, bend left=30] (iTi) to (phi1);
	\draw [-, dashed, bend left=20] (iTi) to (phi2);
	\draw [-, dashed] (iTi) to (phi3);
	\draw [-, dashed, bend right=20] (iTi) to (phi4);
	\draw [-, dashed, bend right=31] (iTi) to (phi5);
	\draw [-, dashed, bend right=38] (iTi) to (phi6);

	\draw [-, ] (phi1) to (phi2);
	\draw [-, ] (phi2) to (phi3);
	\draw [-, ] (phi3) to (phi4);
	\draw [-, ] (phi4) to (phi5);
	\draw [-, ] (phi5) to (phi6);
	\draw [-latex', ] (phi6) to (avgphi);
	\end{tikzpicture}
\end{center}
\noindent Step 3, re-sample $I_i$ on the average transformation $\pmb{avg}_i$, indicated as the next diagram and shown in Figure 7, to get the biased temporary templates ${\hat{T}emplate}_i=I_{i}(\pmb{avg}_i)$ for each $i=1,2,\dots,6$ which are shown in Figure 8. 

\tikzstyle{block} = [rectangle, draw, fill=gray!20, text width=2cm, text centered, rounded corners, minimum height=1cm, node distance=4cm]
\tikzstyle{cloud} = [ellipse, fill=red!10, node distance=4cm, minimum height=2em]
\begin{center}
	\begin{tikzpicture}
	\node [block] (iTi) {$I_{i}$ (initial template)};
	\node [cloud, right of=iTi] (avgphi) {$\pmb{avg}_i$};
	\node [block, right of=avgphi] (TTi) {${\hat{T}emplate}_i$};	
	
	\draw [-, ] (iTi) to (avgphi);
	\draw [-latex', ] (avgphi) to (TTi);
	
	\end{tikzpicture}
\end{center}

\begin{figure}[H]
	\caption{Average transformation --- $\pmb{avg}_i=avg(\pmb{\phi}_{ij=1,\dots,6})$}
	\subfigure[$\pmb{avg}_1$]{\includegraphics[width=3.1cm,height=3.1cm]{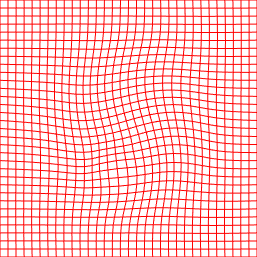}}
	\subfigure[$\pmb{avg}_2$]{\includegraphics[width=3.1cm,height=3.1cm]{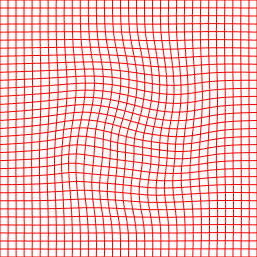}}
	\subfigure[$\pmb{avg}_3$]{\includegraphics[width=3.1cm,height=3.1cm]{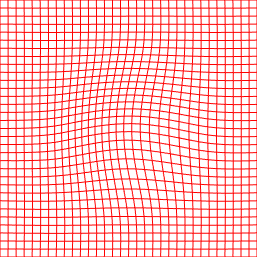}}
	\subfigure[$\pmb{avg}_4$]{\includegraphics[width=3.1cm,height=3.1cm]{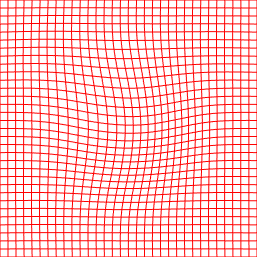}}
	\subfigure[$\pmb{avg}_5$]{\includegraphics[width=3.1cm,height=3.1cm]{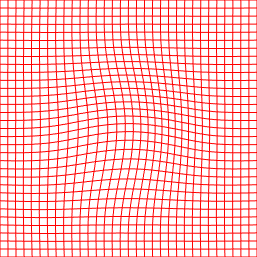}}
	\subfigure[$\pmb{avg}_6$]{\includegraphics[width=3.1cm,height=3.1cm]{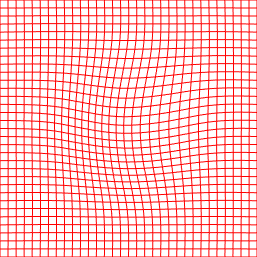}}
\end{figure}

\begin{figure}[H]
	\caption{Biased Temporary Templates --- $\hat{T}emplate_i=I_{i}(\pmb{avg}_i)$}
	\subfigure[$\hat{T}emplate_1$]{\includegraphics[width=3.1cm,height=3.1cm]{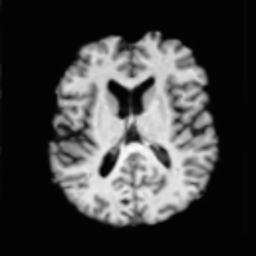}}
	\subfigure[$\hat{T}emplate_2$]{\includegraphics[width=3.1cm,height=3.1cm]{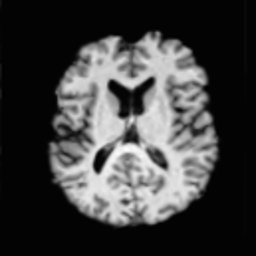}}
	\subfigure[$\hat{T}emplate_3$]{\includegraphics[width=3.1cm,height=3.1cm]{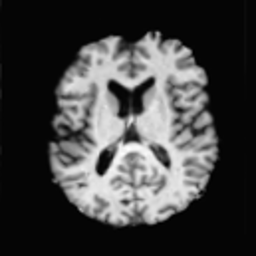}}
	\subfigure[$\hat{T}emplate_4$]{\includegraphics[width=3.1cm,height=3.1cm]{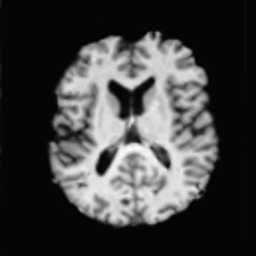}}
	\subfigure[$\hat{T}emplate_5$]{\includegraphics[width=3.1cm,height=3.1cm]{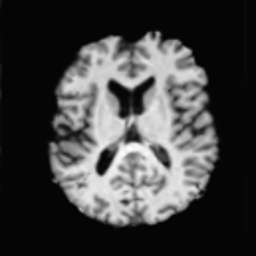}}
	\subfigure[$\hat{T}emplate_6$]{\includegraphics[width=3.1cm,height=3.1cm]{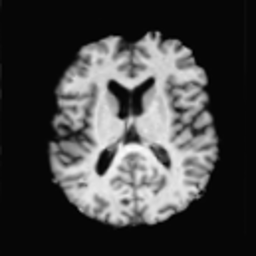}}
\end{figure}

\noindent The SSD between $\hat{I}_i$ with the GT --- $SSD({\hat{T}emplate}_i, I_0)$ and $Error_i = \frac{SSD({\hat{T}emplate}_i, I_0)}{SSD(I_i, I_0)}$ are shown in the following table. $Error_i$ measures the error reduction in temporary template ${\hat{T}emplate}_i$ compared to the image $I_i$.

\vspace{0.2cm}
\begin{minipage}{2\textwidth}
	\begin{tabular}{|c|c|c|c|c|c|c|}
		\hline
		$i$ & $1$ & $2$ & $3$ & $4$ & $5$ & $6$ \\
		\hline
		$SSD(\hat{T}emplate_i, I_0)=(10^{6})*$ & 2.0504 & 3.4073 & 0.4963 & 0.4905 & 0.4835 & 0.5116 \\
		\hline
		$Error_i$ & 0.1084 & 0.1802 & 0.0261 & 0.0250 & 0.0264 & 0.0279  \\
		\hline
	\end{tabular}
\end{minipage}
\vspace{0.2cm}

As it can be seen, ${\hat{T}emplate}_2$ has the highest error $= 0.1802$. At this stage, ${\hat{T}emplate}_4$ is the best with an error 0.0250 (this means $97.5\%$ of the initial $SSD$ between image $I_4$ and GT is reduced by our optimal control registration method). The large difference between $Error_2 = 0.1802$ and $Error_4 = 0.0250$ indicates the existence of bias towards the image that is used as an initial template. In order to quantify the bias, we calculate the sample mean and standard deviation of the six $SSD({\hat{T}emplate}_i, I_0)$'s: Sample Mean $= 1.2400*(10^{6})$, Sample Standard Deviation $= 1.2306*(10^{6})$. So, $SSD(\hat{T}emplate_6,I_0)$ is the closest to the Sample Mean. But the large standard deviation of this sample disqualifies any one of these templates as unbiased. 

This leads to Step-4: repeat Step-1 to Step-3 on biased temporary templates --- $\hat{T}emplate_i$ to reduce their bias. So we get a new group of average transformation $\pmb{Avg}_i$ as shown in Figure 9 and the re-sampled images are the Unbiased Templates --- $Template_i=\hat{T}emplate_{i}(\pmb{Avg}_i)$ as shown in Figure 10.

\begin{figure}[H]
	\caption{Average transformation --- $\pmb{Avg}_i$}
	\subfigure[$\pmb{Avg}_1$]{\includegraphics[width=3.1cm,height=3.1cm]{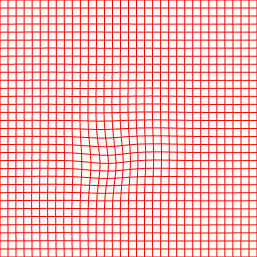}}
	\subfigure[$\pmb{Avg}_2$]{\includegraphics[width=3.1cm,height=3.1cm]{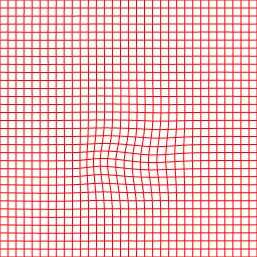}}
	\subfigure[$\pmb{Avg}_3$]{\includegraphics[width=3.1cm,height=3.1cm]{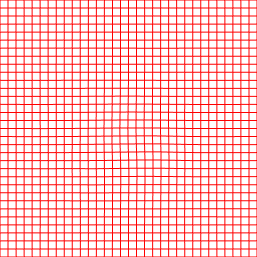}}
	\subfigure[$\pmb{Avg}_4$]{\includegraphics[width=3.1cm,height=3.1cm]{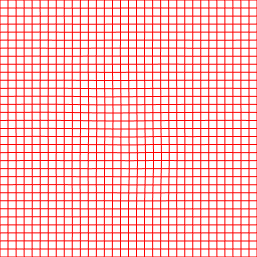}}
	\subfigure[$\pmb{Avg}_5$]{\includegraphics[width=3.1cm,height=3.1cm]{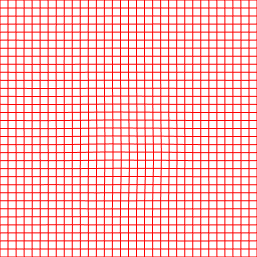}}
	\subfigure[$\pmb{Avg}_6$]{\includegraphics[width=3.1cm,height=3.1cm]{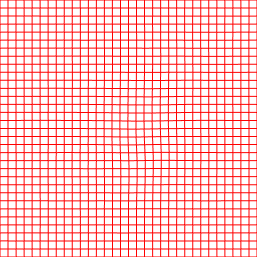}}
\end{figure}

\begin{figure}[H]
	\caption{Unbiased Templates --- $Template_i=\hat{T}emplate_{i}(\pmb{Avg}_i)$}
	\subfigure[$Template_1$]{\includegraphics[width=3.1cm,height=3.1cm]{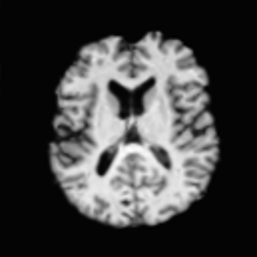}}
	\subfigure[$Template_2$]{\includegraphics[width=3.1cm,height=3.1cm]{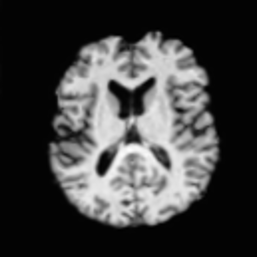}}
	\subfigure[$Template_3$]{\includegraphics[width=3.1cm,height=3.1cm]{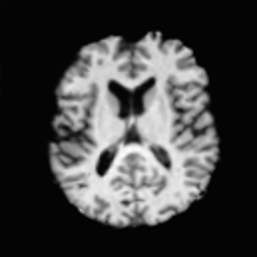}}
	\subfigure[$Template_4$]{\includegraphics[width=3.1cm,height=3.1cm]{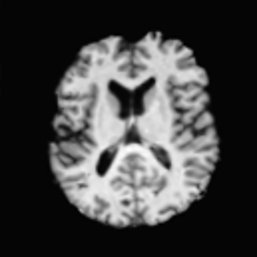}}
	\subfigure[$Template_5$]{\includegraphics[width=3.1cm,height=3.1cm]{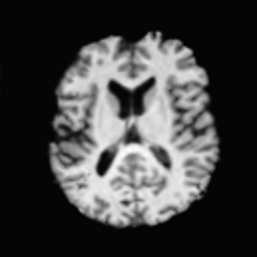}}
	\subfigure[$Template_6$]{\includegraphics[width=3.1cm,height=3.1cm]{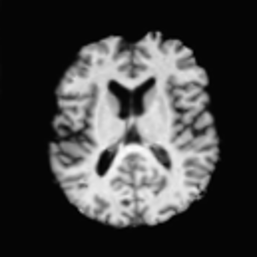}}
\end{figure}

\noindent The SSD between $Template_i$ with the GT --- $SSD({Template}_i, I_0)$ and $Error_i = \frac{SSD({Template}_i, I_0)}{SSD(I_i, I_0)}$ are shown in the following table which measures the error reduction in temporary template ${Template}_i$ compared to the image $I_i$.

\vspace{0.2cm}
\begin{minipage}{2\textwidth}
	\begin{tabular}{|c|c|c|c|c|c|c|}
		\hline
		$i$ & $1$ & $2$ & $3$ & $4$ & $5$ & $6$ \\
		\hline
		$SSD(Template_i, I_0)=(10^{5})*$ & 6.7786 & 6.6041 & 5.5583 & 5.6649 & 5.7469 & 5.8088  \\
		\hline
		$Error_i$ & 0.0358 & 0.0349 & 0.0292 & 0.0289 & 0.0314 & 0.0317  \\
		\hline
	\end{tabular}
\end{minipage}
\vspace{0.2cm}

\noindent The statistics are significantly improved: New Sample Mean $= 6.0269*(10^{5})$; New Sample Standard Deviation $= 5.2436*(10^{4})$. $Template_6$ is closest to the New Sample Mean. The New Sample Standard Deviation $= 5.2436*(10^{4})$ is now only $4.3566\%$ of the previous Sample Standard Deviation $= 1.2306*(10^{6})$. This means, the effectiveness of repeating Step-1 to Step-3 on the biased temporary templates has greatly reduced the bias of biased tmporary template $\{\hat{T}emplate_i\}$. Hence, we can take any of the new templates $Template_{i}$ as an unbiased template. To check that, we may register each of $Template_{i}$ to $I_0$ to get six register transformations and all of them are expected to be close to the identity map $\pmb{Id}$ as the results are shown in Figure 11. 

\begin{figure}[H]
	\caption{Average transformation --- $\pmb{\hat{I}d}_i$}
	\subfigure[$\pmb{\hat{I}d}_1$]{\includegraphics[width=3.1cm,height=3.1cm]{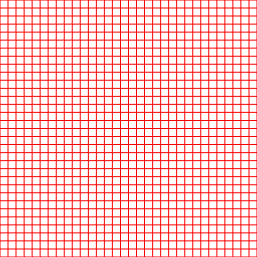}}
	\subfigure[$\pmb{\hat{I}d}_2$]{\includegraphics[width=3.1cm,height=3.1cm]{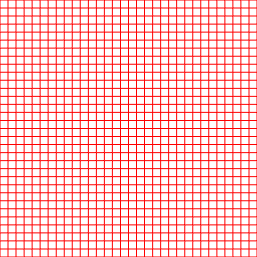}}
	\subfigure[$\pmb{\hat{I}d}_3$]{\includegraphics[width=3.1cm,height=3.1cm]{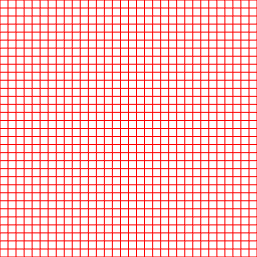}}
	\subfigure[$\pmb{\hat{I}d}_4$]{\includegraphics[width=3.1cm,height=3.1cm]{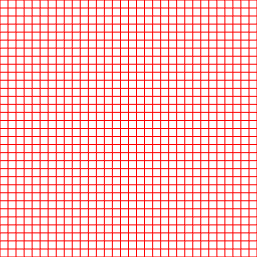}}
	\subfigure[$\pmb{\hat{I}d}_5$]{\includegraphics[width=3.1cm,height=3.1cm]{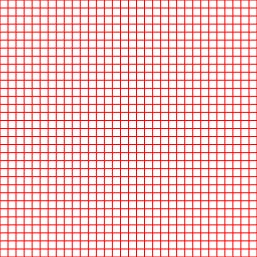}}
	\subfigure[$\pmb{\hat{I}d}_6$]{\includegraphics[width=3.1cm,height=3.1cm]{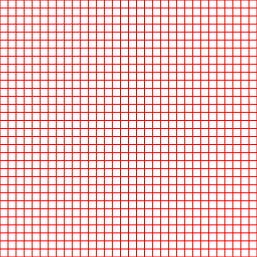}}
\end{figure}

\noindent And the behaviors of their Jacobian determinant and curl are shown in the following table.

\vspace{0.2cm}
\begin{minipage}{2\textwidth}
	\begin{tabular}{|c|c|c|c|c|c|c|}
		\hline
		$i$ & $1$ & $2$ & $3$ & $4$ & $5$ & $6$ \\
		\hline
		max $J(\pmb{\hat{I}d}_i)$ & 1.0000 & 1.0000 & 1.0000 & 1.0000 & 1.0000 & 1.0000  \\
		\hline
		min $J(\pmb{\hat{I}d}_i)$ & 0.9999 & 0.9999 & 0.9999 & 0.9999 & 0.9999 & 0.9999  \\
		\hline
		max $\text{curl}(\pmb{\hat{I}d}_i)=10^{-14}*$ & 0.3553 & 0.3553 & 0.1776 & 0.1776 & 0.3553 & 0.3553  \\
		\hline
		min $\text{curl}(\pmb{\hat{I}d}_i)=10^{-14}*$ & -0.3553 & -0.3553 & -0.1776 & -0.1776 & -0.3553 & -0.3553  \\
		\hline
	\end{tabular}
\end{minipage}
\vspace{0.2cm}

\subsubsection*{Example 4: Computation Friendly Approach}
In Example 3, the crucial part is Step-3 for our general approach of constructing an unbiased template, which requires to find all $\hat{T}emplate_i$ for all $i=1,\dots,6$ and it is not a cheap task to do in terms of computational costs. In this example, we construct a correction transformation to reduce the bias of any initial template $I_i$, by using an approximation to composition of small deformations, which ``allows" us to do a similar job without finding all $\hat{T}emplate_i$. This can be done under the assumption that ground-truth Jacobian determinant and curl-vector are known.

Let's take one of the $I_i$ as the initial template, then we perform Step-1 and Step-2 as they were in Example 3, to get an biased temporary template $\hat{T}emplate_i$. To see the process is independent of choosing initial template, we will show the results of all six image $I_i$ as the initial template individually.

\subsubsection*{Correction transformation $\pmb{H}_{i}$'s}
We have derived a mathematical formula to generate correction transformations $\pmb{H}_{i}$'s for each of $\hat{T}emplate_i$ where $i=1,\dots, 6$. Let $\{I_{i}\}$ be the given images and $\{{\hat{T}emplate}_i\}$ be the temporary templates by above procedures. As it can be seen in Example 3, using $I_1$ and $I_2$ to form the temporary templates would give strongly biased results. Therefore, in order to construct unbiased template, a correction transformation $\pmb{H}_{i}$ needs to be introduced which depends on the initial template $I_{i}$. 

To construct $\pmb{H}_{i}$(arrowed solid-line on the diagram) for a fixed $i$:  Firstly, applying our image registration method to find registration transformations $\hat{\pmb{\phi}}_{ij}: {\hat{T}emplate}_i \longrightarrow I_{j}$, $\forall$ $j=1,2...,N$ (arrowed dash-line on the diagram). Let consider the following diagram:
\tikzstyle{block} = [rectangle, draw, fill=gray!20, text width=1.5cm, text centered, rounded corners, minimum height=1cm, node distance=2.5cm]
\tikzstyle{cloud} = [ellipse, fill=red!10, node distance=2.5cm, minimum height=2em]
\begin{center}
	\begin{tikzpicture}
	\node [block] (I1) {$I_{1}$};
	\node [block, right of=I1] (I2) {$I_{2}$};
	\node [block, right of=I2] (I3) {$I_{3}$};
	\node [block, right of=I3] (I4) {$I_{4}$};
	\node [block, right of=I4] (I5) {$I_{5}$};
	\node [block, right of=I5] (I6) {$I_{6}$};
	
	\node [cloud, below of=I1] (phi1) {$\pmb{\hat{\phi}}_{i1}$};
	\node [cloud, right of=phi1, below of=I1] (phi2) {$\pmb{\hat{\phi}}_{i2}$};
	\node [cloud, right of=phi2, below of=I2] (phi3) {$\pmb{\hat{\phi}}_{i3}$};
	\node [cloud, right of=phi3, below of=I3] (phi4) {$\pmb{\hat{\phi}}_{i4}$};
	\node [cloud, right of=phi4, below of=I4] (phi5) {$\pmb{\hat{\phi}}_{i5}$};
	\node [cloud, right of=phi5, below of=I5] (phi6) {$\pmb{\hat{\phi}}_{i6}$};
	
	\node [block, below of=phi3] (TTi) {$\hat{T}emplate_{i}$};
	
	\node [cloud, below left of=TTi] (Hi) {$\pmb{H}_{i}$};
	\node [block, below of=phi1] (Ti) {$Template_{i}$};

	\draw [-latex', dashed, bend left=10] (phi1) to (I1);
	\draw [-latex', dashed, bend left=6] (phi2) to (I2);
	\draw [-latex', dashed] (phi3) to (I3);
	\draw [-latex', dashed, bend right=3] (phi4) to (I4);
	\draw [-latex', dashed, bend right=5] (phi5) to (I5);
	\draw [-latex', dashed, bend right=8] (phi6) to (I6);
	
	\draw [-, dashed, bend left=30] (TTi) to (phi1);
	\draw [-, dashed, bend left=20] (TTi) to (phi2);
	\draw [-, dashed] (TTi) to (phi3);
	\draw [-, dashed, bend right=20] (TTi) to (phi4);
	\draw [-, dashed, bend right=30] (TTi) to (phi5);
	\draw [-, dashed, bend right=40] (TTi) to (phi6);
	
	\draw [-latex', bend right=25] (Hi) to (TTi);
	\draw [-, bend right=25] (Ti) to (Hi);
	\end{tikzpicture}
\end{center}

\noindent The construction of biased temporary templates $\hat{T}emplate_{i}$ are actually expected to be the unbiased template $Template_{i}$ in Step-1 and Step-2 of Example 3. Due to computational limitations, they did not reduce the bias enough, so we got only a biased temporary template $\hat{T}emplate_{i}$. However, the original bias of the initial template $I_i$ has been reduced to a level that makes the temporary templates $\hat{T}emplate_{i}$ closer to $I_0$ than the initial template $I_i$. We would like to ask, does it exist a small deformation $\pmb{H}_{i}(\pmb{x})$ that deforms the unbiased template $Template_{i}$ to $\hat{T}emplate_{i}$? If it exists, how to find it?

\noindent Secondly, according to our definition of unbiased template, if $\pmb{H}_{i}$ exists, then it must satisfy the following equations:
\begin{equation}\label{H1}
\left\{	\begin{aligned}
\frac{\sum_{j=1}^{N}J(\hat{\pmb{\phi}}_{ij} \circ \pmb{H}_{i}(\pmb{x}))}{N}&=1\\
\sum_{j=1}^{N}\text{curl}(\hat{\pmb{\phi}}_{ij} \circ \pmb{H}_{i}(\pmb{x}))&=0
\end{aligned}\right.
\end{equation}
This means, if there exists an unbiased template which should be almost identical to GT --- $I_0$, then the each compositions of $\hat{\pmb{\phi}}_{ij} \circ \pmb{H}_{i}$ should behaves very close to the inverse transformation of $\pmb{D}_{i}$, where $\pmb{D}_{i}$ were shown in Figure 5. Therefore the average of $J(\hat{\pmb{\phi}}_{ij} \circ \pmb{H}_{i}(\pmb{x}))$ should be $1$ and the average $\text{curl}(\hat{\pmb{\phi}}_{ij} \circ \pmb{H}_{i}(\pmb{x}))$ should be $0$ for each $i$.

By property $J(\pmb{T}_2 \circ \pmb{T}_1)=J(\pmb{T}_1)J(\pmb{T}_2)$, we know 
\begin{equation}\label{H2}
\frac{\sum_{j=1}^{N}J(\hat{\pmb{\phi}}_{ij} \circ \pmb{H}_{i}(\pmb{x}))}{N}=J(\pmb{H}_{i}(\pmb{x}))\frac{\sum_{j=1}^{N}J(\hat{\pmb{\phi}}_{ij}(\pmb{x}))}{N}=1 \hspace{1cm} \Longrightarrow \hspace{1cm} J(\pmb{H}_{i}(\pmb{x}))=\frac{N}{\sum_{j=1}^{N}J(\hat{\pmb{\phi}}_{ij}(\pmb{x}))}
\end{equation}

$\pmb{H}_{i}(\pmb{x})$ is small deformation and it can be represented as $\pmb{H}_{i}(\pmb{x})=\pmb{x}+\pmb{u}_{i}(\pmb{x})$, therefore, we have
\begin{equation}\label{H3}
 \text{curl}(\pmb{H}_{i}(\pmb{x}))=\text{curl}(\pmb{u}_{i}(\pmb{x}))
\end{equation}
And composition of two small deformations, let $\pmb{T}_1(\pmb{x})=\pmb{x}+\pmb{u}_{1}(\pmb{x})$ and $\pmb{T}_2(\pmb{x})=\pmb{x}+\pmb{u}_{2}(\pmb{x})$ where $\pmb{u}_{i}(\pmb{x})$ are displacement fields, can be approximated \cite{Joshi} by

\begin{equation*}
\pmb{T}_2 \circ \pmb{T}_1 (\pmb{x}) \approx \pmb{x}+\pmb{u}_{1}(\pmb{x})+\pmb{u}_{2}(\pmb{x}) =\pmb{x}+\pmb{u}_{2}(\pmb{x})+\pmb{u}_{1}(\pmb{x}) =\pmb{T}_2 (\pmb{x})+\pmb{u}_{1}(\pmb{x})
\end{equation*}
this leads to 
\begin{equation*}
\pmb{H}_{i} \circ \hat{\pmb{\phi}}_{ij}(\pmb{x})\approx \hat{\pmb{\phi}}_{ij}(\pmb{x}) + \pmb{u}_{i}(\pmb{x})
\end{equation*}
Therefore, by (\ref{H3}), we have
\begin{equation}\label{H4}
\text{curl}(\pmb{H}_{i} \circ \hat{\pmb{\phi}}_{ij}(\pmb{x})) \approx \text{curl}(\hat{\pmb{\phi}}_{ij})+ \text{curl}(\pmb{u}_{i}(\pmb{x})) =\text{curl}(\hat{\pmb{\phi}}_{ij})+ \text{curl}(\pmb{H}_{i}(\pmb{x}))
\end{equation}

\noindent Now, by (\ref{H2}) and (\ref{H4}), the approximations of Jacobian determinant and curl of $\pmb{H}_i$ in (\ref{H1}), denoted as $J(\hat{\pmb{H}}_{i}(\pmb{x}))$ and $\text{curl}(\hat{\pmb{H}}_{i}(\pmb{x}))$, respectively, which are in the form of 
\begin{equation}\label{H5}
\left\{
\begin{aligned}
J(\hat{\pmb{H}}_{i}(\pmb{x}))&=\frac{N}{\sum_{j=1}^{N}J(\hat{\pmb{\phi}}_{ij}(\pmb{x}))}\\
\text{curl}(\hat{\pmb{H}}_{i}(\pmb{x}))&= \frac{-\sum_{j=1}^{N}\text{curl}(\hat{\pmb{\phi}}_{ij}(\pmb{x}))}{N}
\end{aligned}\right.
\end{equation}

\noindent Finally, we find $\hat{\pmb{H}}_{i}$ by our Variational Principle. The approximated transformations $\hat{\pmb{H}}_{i}$'s are shown in Figure 12.
\begin{figure}[H]
	\caption{Correction Transformation, $\hat{\pmb{H}}_{1-6}$}
	\subfigure[$\hat{\pmb{H}}_{1}$]{\includegraphics[width=3.1cm,height=3.1cm]{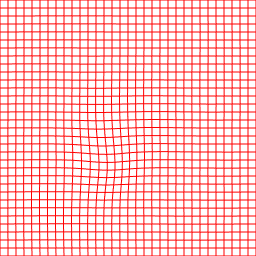}}
	\subfigure[$\hat{\pmb{H}}_{2}$]{\includegraphics[width=3.1cm,height=3.1cm]{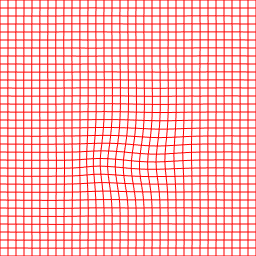}}
	\subfigure[$\hat{\pmb{H}}_{3}$]{\includegraphics[width=3.1cm,height=3.1cm]{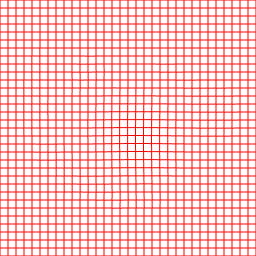}}
	\subfigure[$\hat{\pmb{H}}_{4}$]{\includegraphics[width=3.1cm,height=3.1cm]{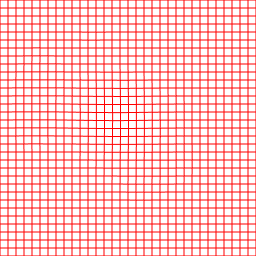}}
	\subfigure[$\hat{\pmb{H}}_{5}$]{\includegraphics[width=3.1cm,height=3.1cm]{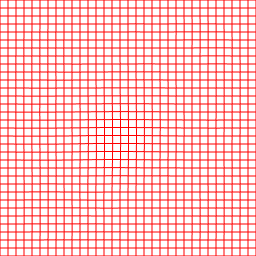}}
	\subfigure[$\hat{\pmb{H}}_{6}$]{\includegraphics[width=3.1cm,height=3.1cm]{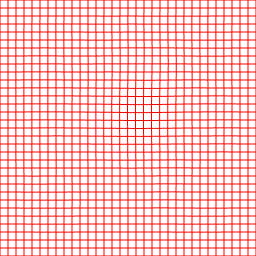}}
\end{figure}

\noindent After re-sampling each $\hat{T}emplate_{i}$ on transformation $\hat{\pmb{H}}_{i}$'s, we obtain a set of six new templates $Template_{i}$ as shown in Figure 13, which are much more uniform now. 
\begin{figure}[H]
	\caption{Unbiased Templates --- $Template_{i}$}
	\subfigure[$Template_{1}$]{\includegraphics[width=3.1cm,height=3.1cm]{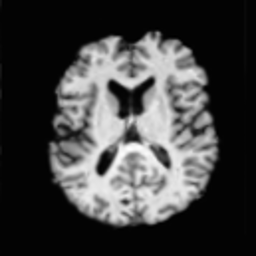}}
	\subfigure[$Template_{2}$]{\includegraphics[width=3.1cm,height=3.1cm]{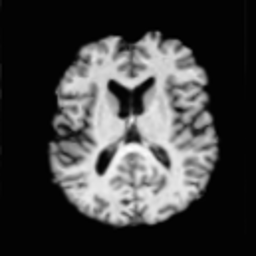}}
	\subfigure[$Template_{3}$]{\includegraphics[width=3.1cm,height=3.1cm]{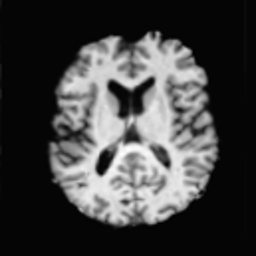}}
	\subfigure[$Template_{4}$]{\includegraphics[width=3.1cm,height=3.1cm]{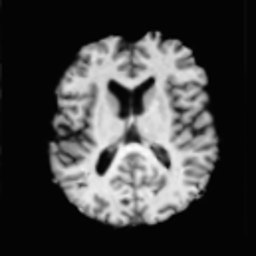}}
	\subfigure[$Template_{5}$]{\includegraphics[width=3.1cm,height=3.1cm]{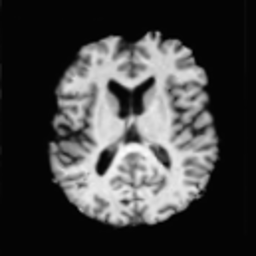}}
	\subfigure[$Template_{6}$]{\includegraphics[width=3.1cm,height=3.1cm]{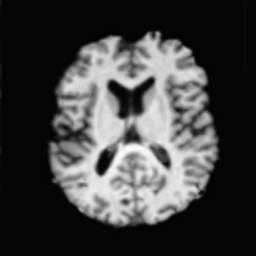}}
\end{figure}

\noindent Similarly as the results from Example 3, the highest error is $0.0372$, downed from $0.1802$. The $SSD$ between new templates $Template_i$'s and $I_0$ are reduced by an order of $10$ as $SSD(Template_i, I_0)$ and $Error_i = \frac{SSD(Template_i, I_0)}{SSD(I_i, I_0)}$ shown in the table. It measures the relative error in the new template $Template_i$ compared to the image $I_i$. 
 
\vspace{0.2cm}
\begin{minipage}{2\textwidth}
	\begin{tabular}{|c|c|c|c|c|c|c|}
		\hline
		$i$ & $1$ & $2$ & $3$ & $4$ & $5$ & $6$ \\
		\hline
		$SSD(Template_i, I_0)=(10^{5})*$ & 6.6972 & 7.0447 & 5.0143 & 4.9911 & 5.0290 & 5.2408  \\
		\hline
		$Error_i$ & 0.0354 & 0.0372 & 0.0264 & 0.0254 & 0.0274 & 0.0286 \\
		\hline
	\end{tabular}
\end{minipage}
\vspace{0.2cm}

\noindent The statistics are also significantly improved just like in the results in Example 3: New Sample Mean $= 5.6695*(10^{5})$; New Sample Standard Deviation $= 9.4137*(10^{4})$. $Template_6$ is closest to the New Sample Mean. The New Sample Standard Deviation $= 9.4137*(10^{4})$ is now only $7.65\%$ of the previous Sample Standard Deviation $= 1.2306*(10^{6})$. This means, with the help of approximated correction transformations $\hat{\pmb{H}}_{i}$, we have greatly reduced the bias. And $\hat{\pmb{H}}_{i}$ has been a good approximation to $\pmb{H}_{i}$. To check that, we registered $Template_{i}$ to $I_0$ for each $i$, denoted as $\pmb{\hat{I}d}_i$ which are shown in next Figure 14. They are expected to be close to the identity map $\pmb{Id}$. Therefore, we can take any of the new templates $Template_{i}$ as an unbiased template.

\begin{figure}[H]
	\caption{$\pmb{\hat{I}d}_i$}
	\subfigure[$\pmb{\hat{I}d}_1$]{\includegraphics[width=3.1cm,height=3.1cm]{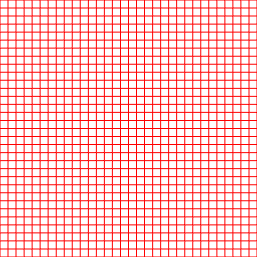}}
	\subfigure[$\pmb{\hat{I}d}_2$]{\includegraphics[width=3.1cm,height=3.1cm]{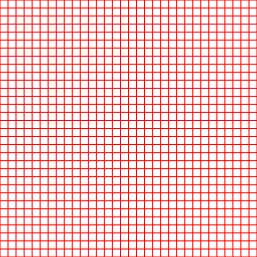}}
	\subfigure[$\pmb{\hat{I}d}_3$]{\includegraphics[width=3.1cm,height=3.1cm]{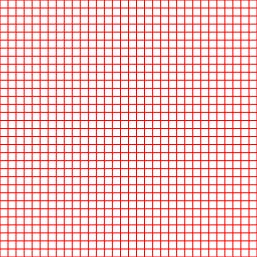}}
	\subfigure[$\pmb{\hat{I}d}_4$]{\includegraphics[width=3.1cm,height=3.1cm]{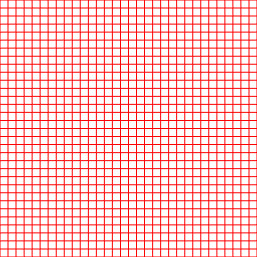}}
	\subfigure[$\pmb{\hat{I}d}_5$]{\includegraphics[width=3.1cm,height=3.1cm]{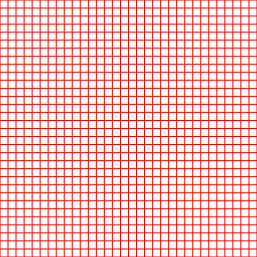}}
	\subfigure[$\pmb{\hat{I}d}_6$]{\includegraphics[width=3.1cm,height=3.1cm]{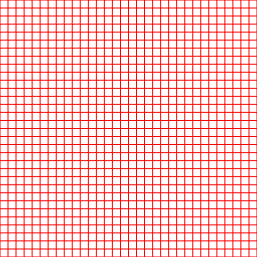}}
\end{figure}

\vspace{0.2cm}
\begin{minipage}{2\textwidth}
		\begin{tabular}{|c|c|c|c|c|c|c|}
			\hline
			$i$ & $1$ & $2$ & $3$ & $4$ & $5$ & $6$ \\
			\hline
			max $J(\pmb{\hat{I}d}_i)$ & 1.0000 & 1.0000 & 1.0000 & 1.0000 & 1.0000 & 1.0000  \\
			\hline
			min $J(\pmb{\hat{I}d}_i)$ & 0.9999 & 0.9999 & 0.9999 & 0.9999 & 0.9999 & 0.9999  \\
			\hline
			max $\text{curl}(\pmb{\hat{I}d}_i)=10^{-14}*$ & 0.3553 & 0.3553 & 0.1776 & 0.1776 & 0.1776 & 0.1776  \\
			\hline
			min $\text{curl}(\pmb{\hat{I}d}_i)=10^{-14}*$ & -0.3553 & -0.3553 & -0.1776 & -0.1776 & -0.1776 & -0.1776  \\
			\hline
		\end{tabular}
\end{minipage}
\vspace{0.2cm}

\section{Optimal Control Method for Non-rigid Image Registration}
In this section, we describe the image registration method that we used to calculate the examples. Image registration is the process of establishing pixel (voxel) correspondence between two or more images so that certain similarity measure is maximized, or a dis-similarity measure ($DS$) is minimized. The images can be taken from the same individual at different posts and different times; or from different individuals. The images can also be in different modalities.

Over last decades, many sophisticated methods have been developed. Most of these methods regularize the ill-posed registration problem by various penalty terms that are either based on physical models or geometric considerations. The cost to be minimized is the sum of $D$S and the regularizing term Reg: $C = DS + \alpha Reg$, where $\alpha$ is a parameter controlling the trade-off between the data term $DS$ and the regularizing term $Reg$. There are two issues with this framework: 

\begin{enumerate}[(1)]
	\item the addition of the second term distorted the problem; For instance, the regularizing term changed the optimal transportation flow to be non-gradient flow.
	
	\item The determination of the value of $\alpha$ is more like an art than science. If it is too large, the registration transformation becomes too smooth; while if it is too small the computation becomes unstable.
\end{enumerate}

Some methods introduce regularization implicitly. For instance, the splines-based method minimizes $DS$ on a coarse grid (say 8 by 8 pixels are represented in a grid cell) whose nodes are the control points for splines. After the new locations of the control points are determined, locations of the remaining pixels are interpolated by the splines formulas.

In \cite{HsiaoImgReg,Liao} we proposed an optimal control approach that has no explicit penalty terms (and hence no parameters). Instead, we minimize $DS$ subject to the constraints $\mathcal{L}[\pmb{u}] = \pmb{F}$ with respect to the control function $\pmb{F}$. The registration transformation T is iteratively determined: at the $k+1$ step, $\pmb{T}^{k+1}(\pmb{x}) = \pmb{T}^{k} (\pmb{x} + \pmb{u}^{k+1})$. $\mathcal{L}$ is a partial differential operator providing regularity. For instance, in 2D, we can take $\text{div}(\pmb{u}) = F_1$, and $\text{curl}(\pmb{u}) = F_2$. Or we can simply take $\mathrm{\Delta} \pmb{u} = \pmb{F}$. In the version described in \cite{HsiaoImgReg}, there is a mechanism that keeps the Jacobian determinant positive, which in turn assures that $\pmb{T}$ is a diffeomorphism (invertible and smooth).

The optimal control approach is inspired by the success of fluid control algorithms. The variational problem is solved by the gradient decent method. As indicated in the survey paper \cite{Sotiras}, one could add a penalty term to $DS$ for specific application with a small parameter value $\alpha$ in order to use prior knowledge. We now present a teapot example.

\subsubsection*{Example 5: Registration of 3D Teapot Images by Optimal Control Method} 

We now describe the volume image $I_{0}$ and the twisted image $I_{t}$ in Step-1. In Step-2, we register $I_{t}$ to $I_{0}$, and out put the registration transformation and the deformed image $I_{t}$. The results show that the twisted image $I_{t}$ is deformed back to the $I_{0}$ with high accuracy.

Step-1: A teapot is rotated about its vertical axis passing through the tip of its lid. Snapshots are taken from a fixed camera at $5^{\circ}$ intervals. A total of 72 photos are taken as the teapot completes $360^{\circ}$ rotation and return to the initial position. Each of these photos is re-sampled as a $72 \times 72$ grayscale image $I_{0i}$. These 72 images $I_{0i}$ for $i =1,\dots,72$ are used to form a three dimensional volume image of size $72 \times 72 \times 72$. We refer to this volume image as the Rotating Teapot Image, denoted as $I_{0}$.

\begin{figure}[H]
	\caption{Teapot---$I_0$}
	\subfigure[$0^{\circ}$]{\includegraphics[width=2cm,height=2cm]{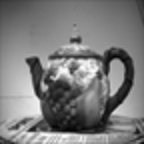}}
	\subfigure[$45^{\circ}$]{\includegraphics[width=2cm,height=2cm]{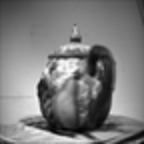}}
	\subfigure[$90^{\circ}$]{\includegraphics[width=2cm,height=2cm]{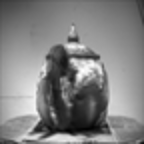}}
	\subfigure[$135^{\circ}$]{\includegraphics[width=2cm,height=2cm]{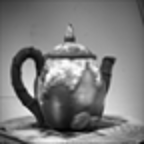}}
	\subfigure[$180^{\circ}$]{\includegraphics[width=2cm,height=2cm]{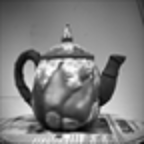}}
	\subfigure[$225^{\circ}$]{\includegraphics[width=2cm,height=2cm]{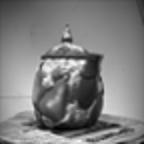}}
	\subfigure[$270^{\circ}$]{\includegraphics[width=2cm,height=2cm]{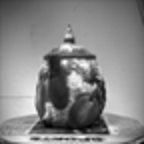}}
	\subfigure[$315^{\circ}$]{\includegraphics[width=2cm,height=2cm]{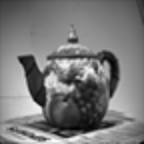}}
	\subfigure[$360^{\circ}$]{\includegraphics[width=2cm,height=2cm]{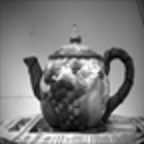}}
\end{figure}

\noindent As the teapot rotates, we deform each slice $I_{0i}$ (of $I_{0}$) to $I_{ti}$ by a rotation $\pmb{T}_i$ as show in Figure 16 that is cut-off near the boundary for $i =1,\dots,72$. 
\begin{figure}[H]
	\caption{$\pmb{T}_i$'s}
	\subfigure[$\pmb{T}_1$]{\includegraphics[width=2cm,height=2cm]{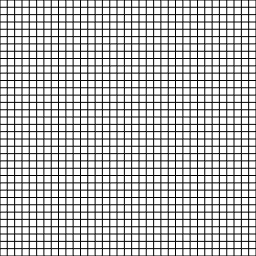}}
	\subfigure[$\pmb{T}_9$]{\includegraphics[width=2cm,height=2cm]{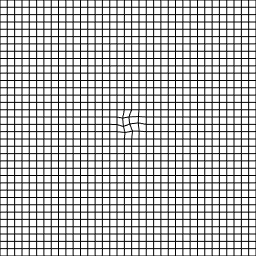}}
	\subfigure[$\pmb{T}_{18}$]{\includegraphics[width=2cm,height=2cm]{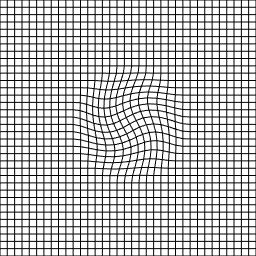}}
	\subfigure[$\pmb{T}_{27}$]{\includegraphics[width=2cm,height=2cm]{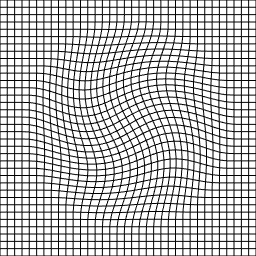}}
	\subfigure[$\pmb{T}_{36}$]{\includegraphics[width=2cm,height=2cm]{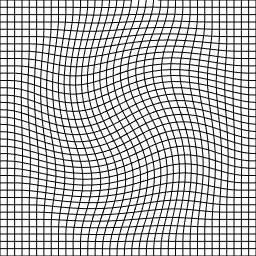}}
	\subfigure[$\pmb{T}_{45}$]{\includegraphics[width=2cm,height=2cm]{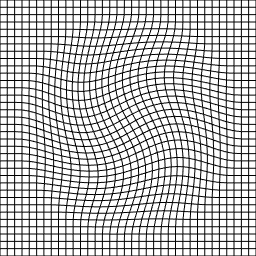}}
	\subfigure[$\pmb{T}_{54}$]{\includegraphics[width=2cm,height=2cm]{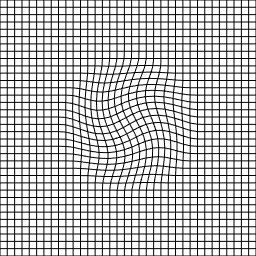}}
	\subfigure[$\pmb{T}_{63}$]{\includegraphics[width=2cm,height=2cm]{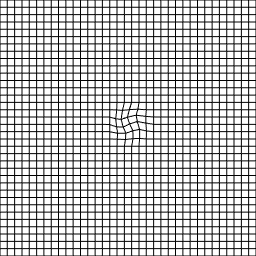}}
	\subfigure[$\pmb{T}_{72}$]{\includegraphics[width=2cm,height=2cm]{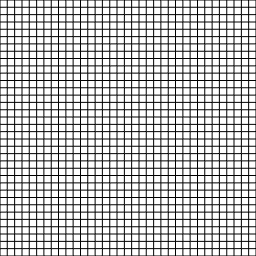}}
\end{figure}

\noindent These 72 twisted $72 \times 72$ images form a $72 \times 72 \times 72$ volume image referred to as the Twisted Teapot, denoted as $I_t$ as shown in Figure 17. 
\begin{figure}[H]
	\caption{Twisted Teapot---$I_t=\pmb{T}(I_0)$}
	\subfigure[$0^{\circ}$]{\includegraphics[width=2cm,height=2cm]{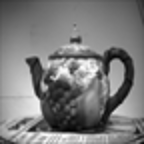}}
	\subfigure[$45^{\circ}$]{\includegraphics[width=2cm,height=2cm]{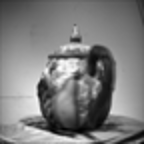}}
	\subfigure[$90^{\circ}$]{\includegraphics[width=2cm,height=2cm]{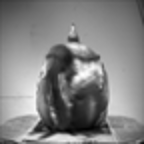}}
	\subfigure[$135^{\circ}$]{\includegraphics[width=2cm,height=2cm]{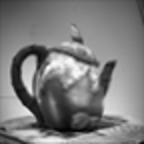}}
	\subfigure[$180^{\circ}$]{\includegraphics[width=2cm,height=2cm]{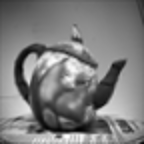}}
	\subfigure[$225^{\circ}$]{\includegraphics[width=2cm,height=2cm]{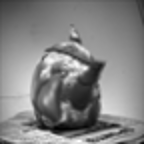}}
	\subfigure[$270^{\circ}$]{\includegraphics[width=2cm,height=2cm]{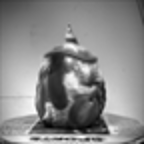}}
	\subfigure[$315^{\circ}$]{\includegraphics[width=2cm,height=2cm]{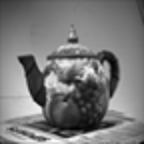}}
	\subfigure[$360^{\circ}$]{\includegraphics[width=2cm,height=2cm]{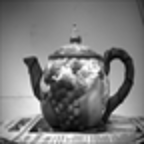}}
\end{figure}
\noindent The deformations $\pmb{T}_i$ for $i =1,\dots,72$ and the Twisted Teapot image $I_t$ can be seen from \href{https://youtu.be/AauiN0329Z8}{Twisted Teapot(CLICK HERE)} 


Step-2:  The registration transformation $\pmb{\phi}$ from $I_0$ to $I_t$ is calculated by minimizing

\begin{equation}
SSD(\pmb{\phi}(\pmb{x}))=\iiint(I_t(\pmb{\phi}(\pmb{x}))-I_0(\pmb{x}))^2d\pmb{x} 
\end{equation}

\noindent The resulting registration deformation $\pmb{\phi}$, restricted to each $i$th slice, is expected to be very close to $\pmb{T}_i$ as shown in Figure 18. 
\begin{figure}[H]
	\caption{$\pmb{\phi}_i$'s}
	\hspace{1cm}
	\subfigure[$\pmb{\phi}_1$]{\includegraphics[width=5cm,height=5cm]{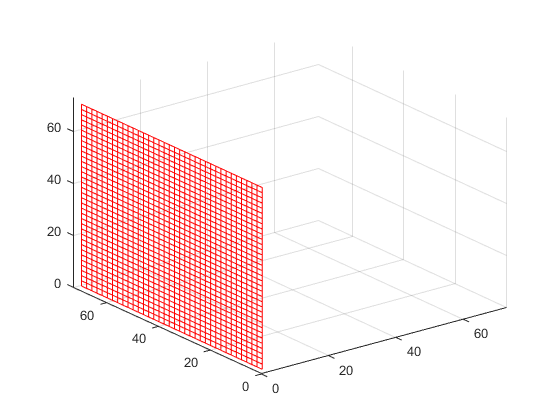}}
	\hspace{1cm}
	\subfigure[$\pmb{\phi}_9$]{\includegraphics[width=5cm,height=5cm]{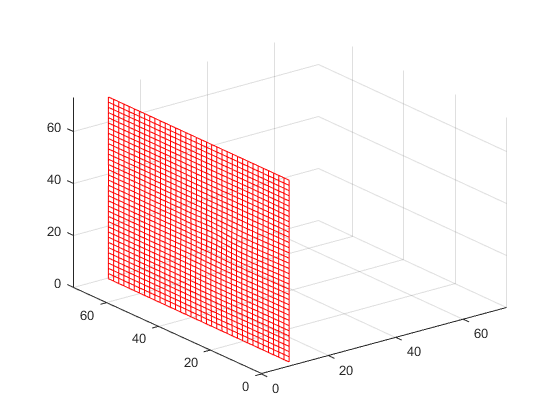}}
	\hspace{1cm}
	\subfigure[$\pmb{\phi}_{18}$]{\includegraphics[width=5cm,height=5cm]{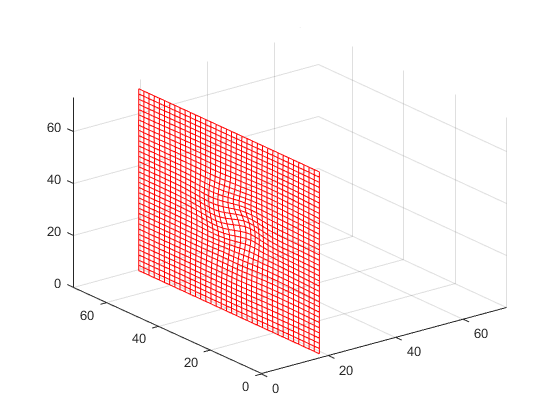}}
	
	\hspace{1cm}
	\subfigure[$\pmb{\phi}_{27}$]{\includegraphics[width=5cm,height=5cm]{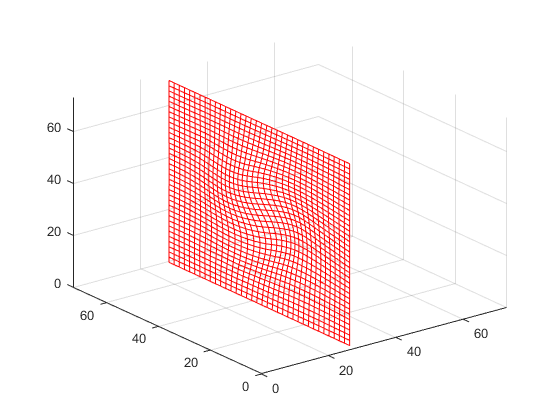}}
	\hspace{1cm}
	\subfigure[$\pmb{\phi}_{36}$]{\includegraphics[width=5cm,height=5cm]{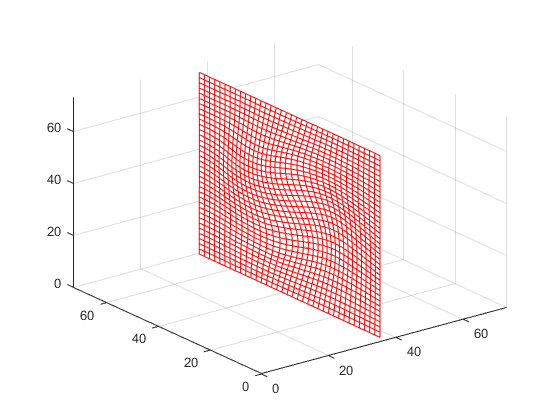}}
	\hspace{1cm}
	\subfigure[$\pmb{\phi}_{45}$]{\includegraphics[width=5cm,height=5cm]{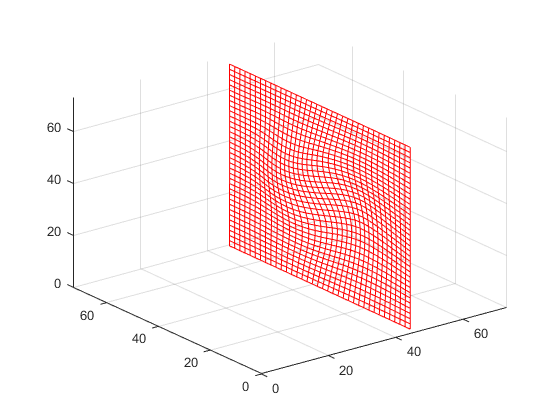}}
	
	\hspace{1cm}
	\subfigure[$\pmb{\phi}_{54}$]{\includegraphics[width=5cm,height=5cm]{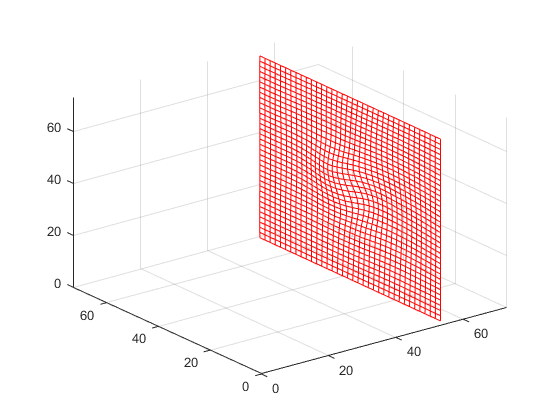}}
	\hspace{1cm}
	\subfigure[$\pmb{\phi}_{63}$]{\includegraphics[width=5cm,height=5cm]{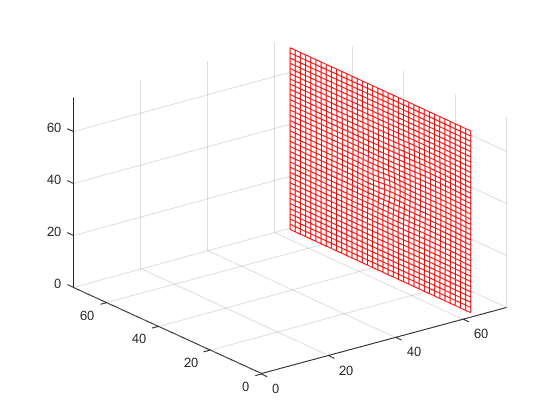}}
	\hspace{1cm}
	\subfigure[$\pmb{\phi}_{72}$]{\includegraphics[width=5cm,height=5cm]{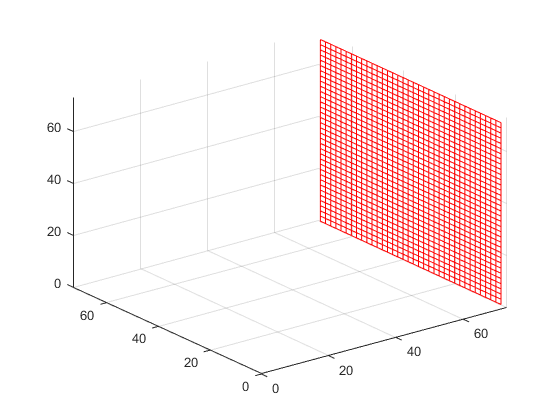}}
\end{figure}
\noindent and the deformed images of the twisted teapot --- $I_t(\pmb{\phi})$ is expected to be close to $I_0$ as it shows on next Figure 19. The total elapsed time for Step-2 is 927.601032 seconds.

\begin{figure}[H]
	\caption{Reversed Twisted Teapot---$I_t(\pmb{\phi})$}
	\subfigure[$0^{\circ}$]{\includegraphics[width=2cm,height=2cm]{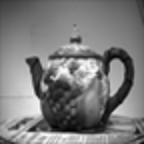}}
	\subfigure[$45^{\circ}$]{\includegraphics[width=2cm,height=2cm]{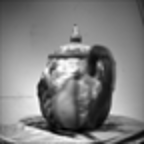}}
	\subfigure[$90^{\circ}$]{\includegraphics[width=2cm,height=2cm]{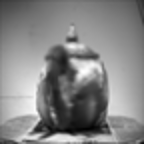}}
	\subfigure[$135^{\circ}$]{\includegraphics[width=2cm,height=2cm]{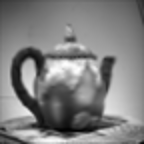}}
	\subfigure[$180^{\circ}$]{\includegraphics[width=2cm,height=2cm]{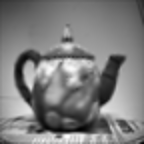}}
	\subfigure[$225^{\circ}$]{\includegraphics[width=2cm,height=2cm]{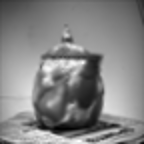}}
	\subfigure[$270^{\circ}$]{\includegraphics[width=2cm,height=2cm]{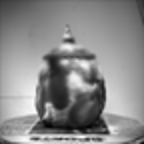}}
	\subfigure[$315^{\circ}$]{\includegraphics[width=2cm,height=2cm]{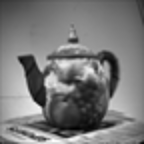}}
	\subfigure[$360^{\circ}$]{\includegraphics[width=2cm,height=2cm]{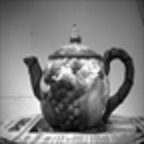}}
\end{figure}

\noindent The registration deformation $\pmb{\phi}$ from Step-2 and the corresponding reversed twisted teapot image $I_t=\pmb{T}(I_0)$ are shown from: \href{https://youtu.be/wIFOM1R8M5I}{Reversed Twisted Teapot(CLICK HERE)}. 
From these visualization files, we conclude that our optimal control method correctly recovered the ground truth deformations $\pmb{T}_i$ for $i =1,\dots,72$ and the grand truth images $I_{0i}$ for $i =1,\dots,72$, as expected.

\section{Conclusion}
In this paper, a set of new computational techniques is described. They are based on both the Jacobian determinant and the curl-vector. Specifically, The Variational Principle is used to calculate a transformation with prescribed Jacobian determinant and curl-vector. A new method of averaging deformations based on averaging the Jacobian determinants and the curl-vectors is used to construct an unbiased template from the members of a set of images. We will further refine the techniques and generalize them to three dimensional real world images.

%

\end{document}